\documentclass[pra,aps,twocolumn,10pt,showpacs,groupedaddress,superscriptaddress,floatfix]{revtex4-1}
\usepackage{amsmath,amsfonts,amssymb,graphics,graphicx,epsfig,color,times}
\usepackage[utf8x]{inputenc}
\usepackage{color}
\usepackage{bbm} 
\usepackage{subfigure}
\usepackage{hyperref}
\usepackage{mathrsfs}
\usepackage{verbatim}
\begin{document}
\title{Extracting Dynamical Green's Function of Ultracold Quantum Gases via Electromagnetically Induced Transparency}
\author{H. H. Jen}
\affiliation{Department of Physics, Technische Universit\"{a}t Kaiserslautern, Kaiserslautern, Germany}
\author{Daw-Wei Wang}
\affiliation{Physics Department, National Tsing Hua University, Hsinchu, Taiwan, R. O. C.}
\affiliation{Physics Division, National Center for Theoretical Sciences, Hsinchu, Taiwan, R. O. C.}

\newcommand{\p}{\mathbf{p}}
\renewcommand{\r}{\mathbf{r}}
\renewcommand{\k}{\mathbf{k}}
\newcommand{\q}{\mathbf{q}}
\date{\today}
\begin{abstract}
The essential quantum many-body physics of an ultracold quantum gas relies on the single-particle Green's functions.\ We demonstrate that it can be extracted by the spectrum of electromagnetically induced transparency (EIT).\ The single-particle Green's function can be reconstructed by the measurements of frequency moments in EIT spectroscopy.\ This optical measurement provides an efficient and nondestructive method to reveal the many-body properties, and we propose an experimental setup to realize it.\ Finite temperature and finite size effects are discussed, and we demonstrate the reconstruction steps of Green's function for the examples of three-dimensional Mott-insulator phase and one-dimensional Luttinger liquid.
\end{abstract}
\maketitle
\section{Introduction}

The ultracold quantum gases have been the test bed to investigate the many-body physics \cite{manybody}.  Thanks to the versatile advances in the experiments, the interaction strength \cite{FB}, the optical lattice depth, and the dimensionality can be manipulated to study strongly correlated quantum gases.  To name a few, the typical atomic systems involve Luttinger liquid of one-dimensional (1D) Bose gases \cite{Haldane1, Haldane2, 1D_confinement, 1D, 1D_Bloch,Esslinger, LL}, a three-dimensional (3D) Bose-Mott insulator \cite{BHM, Bloch}, and the superfluid state of two-component Fermi gases \cite{manyparticle}.  Conventional experimental measurements of the atomic many-body systems include the time-of-flight experiment, the noise correlations measurement \cite{noise}, the Bragg scattering spectroscopy \cite{Bragg1,Bragg2,Bragg3,Bragg4,Bragg5}, and the in-situ imaging \cite{Chin,Ho}. Recently, we investigate the spectrum of electromagnetically induced transparency (EIT) \cite{Lukin_rmp, Fleischhauer_rmp} in the strongly correlated atomic systems \cite{Jen} as an alternative and non-destructive method \cite{pairing, Jen} to probe the quantum many-body physics.

The EIT spectrum is shown to be solely determined by the single-particle Green's function of the ground-state atoms.\ Non-trivial many-body effect for the spectrum is predicted \cite{Jen, Jen2} when the atoms are virtually coupled to the low-lying Rydberg states \cite{Rydberg, 30P}.\  The single-particle Green's function is crucial for many-body systems, from which the observable of any single-particle operator, the ground state energy, and the excitation spectrum can be extracted \cite{manyparticle}.\ Similar to the well-known angle-resolved photoemission spectroscopy (ARPES) \cite{Arpes}, where the single-particle Green's function is probed to investigate the electronic structure of the surface in the solids, the EIT spectroscopy accesses the information of it as well.\ What differs is the single-particle Green's function in ARPES is directly measured whereas the EIT spectrum involves an integral of the Green's function and the functional laser parameters \cite{Jen}.\ A recent proposal of probing spin correlation functions by Ramsey interferometry \cite{Knap} provides the other direct way to measure the dynamical Green's function of the many-body systems.\ 

In ultracold atomic systems, matter wave interference patterns are powerful to reveal the many-body physics.\ For example in two-dimensional (2D) Bose-Einstein condensate (BEC) \cite{BKT}, the contrast of interference fringes relates to the spatial single-particle Green's function \cite{BKT2}.\ Therefore, the algebraic decay of spatial correlation in the contrast indicates the quasi-long-range order in the system, which can be fully described by Green's function.\ In addition, the second-order correlation function in the interference pattern of two 1D Bose gases is also manifested to provide the evidence of Luttinger liquid signatures \cite{probe}.\ On the contrary to the destructive interference method, quantum non-demolition (QND) detection of quantum matter \cite{QND} provides a least destructive measurement.\ It can be used to effectively map the quantum correlations of the ultracold atoms into light signals, which is similar to our proposed EIT scheme in this paper.\ The common feature of matter wave interference and EIT spectrum in probing the quantum many-body physics is that the measurements respectively involve the average and integral of the Green's function, making both methods the indirect probes.\ Nevertheless the EIT spectroscopy has the advantage of non-destructive probe with well-controlled laser parameters, which may provide more efficient and precise measurements on the ultracold atoms.

In this paper, we demonstrate that the dynamical single-particle Green's function of the many-body system can be determined by EIT spectroscopy.\ In section II, we briefly describe the EIT susceptibility in terms of the Green's function \cite{Jen}.\ Then we propose an experimental setup to realize the reconstruction of the single-particle Green's function, and also discuss the finite size effect.\ In section III, we investigate finite temperature Green's function, and demonstrate the reconstruction steps by measuring the frequency moments from the information of EIT spectrum.\ We take 3D Mott-insulator and 1D Luttinger Liquid as two examples.\ In the last section, we discuss possible application to other many-body systems and summarize.\ Our results suggest a universal determination on the essential single-particle Green's function of the ultracold quantum gases in parallel to the ARPES in the solids.

\section{Formulation and proposed experimental setup}
\subsection{Formulation of the EIT scheme}

\begin{figure}[t]
\centering\includegraphics[height=4.8cm, width=8cm]{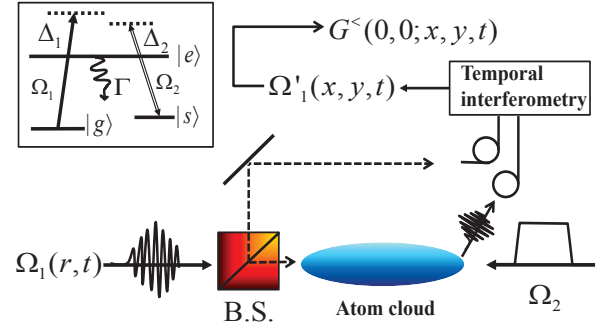}
\caption{(Color online) Schematic setup to extract single-particle Green's function [$G^<(0,0;x,y,t)$] via an EIT experiment with a probe pulse.\ The experiment is conducted with a probe pulse in the ultracold quantum gas with an atomic $\Lambda$-type configuration: The counter-propagating control ($\Omega_{2}$) and the probe [$\Omega_1(\r,t)$] fields in $\hat{z}$ direction couple two hyperfine ground states $|g\rangle$ and $|s\rangle$ with the excited state $|e\rangle$ (detunings are $\Delta_2$ and $\Delta_1$ respectively). $\Gamma$ is the spontaneous decay rate of $|e\rangle$.\ B.S. represents the beam splitter.\ Open circles denote the fiber coupling that guide the reference and transmitted pulses into the temporal interferometry.\ From the interferometry box, we characterize the output probe pulse $\Omega'_1(x,y,t)$.\ $G^<(0,0;x,y,t)$ is then extracted by reconstructing Green's function from the frequency moment calculation in EIT spectroscopy.}%
\label{fig1}
\end{figure}

We consider the conventional EIT setup ($\Lambda$ type scheme) as shown in Fig. \ref{fig1}.\ The probe ($\Omega_1$) and control ($\Omega_2$) fields couple the ground state $|g\rangle$ to the other hyperfine ground state $|s\rangle$ and an excited state $|e\rangle$.  The detunings are $\Delta_1$ and $\Delta_2$ for the probe and control fields respectively.\ They are defined as the differences from the laser central ($\omega_1$) to the atomic transition frequencies.\ In the linear response of the probe field, the electric susceptibility is derived as $\chi(\q,\omega)=\delta\langle\hat{P}(\q,\omega)\rangle/\tilde{\Omega}_{1}(\q,\omega)$ where $\tilde{\Omega}_{1}(\q,\omega)$ is the Fourier transform of slow-varying $\Omega_{1}(\r,t)$, and $\delta\langle\hat{P}(\q,\omega)\rangle$ is the polarization operator calculated by the perturbation of a weak probe field \cite{manyparticle}.

In the momentum-frequency space, the electric susceptibility is obtained as ($\hbar=1$) \cite{Jen}
\begin{align}
\chi(\q,\omega)=&-\frac{d_0}{V}\sum_\k\int_{-\infty}^{\infty}d\tilde{\omega}i\tilde{G}^<(\k,\tilde{\omega})M(\k+\q,\omega-\tilde{\omega}),\label{chi}
\end{align}
where $V$ is the quantization volume, $d_0$ is the dipole moment of the probe field transition, and $\tilde{G}^{<}(\k,\tilde{\omega})$ is the Fourier transform of the Green's function $G^{<}(0,0;\r,t)$ at zero temperature \cite{manyparticle}.\ The functional $M(\k+\q,\omega-\tilde{\omega})$ that includes the laser parameters is 
\begin{align}
M(\k+\q,\omega-\tilde{\omega})&=\frac{\cos^2\phi_{\k+\q}}{\tilde{\omega}-\omega-\epsilon_{-}(\k+\q)}\nonumber\\
&+\frac{\sin^2\phi_{\k+\q}}{\tilde{\omega}-\omega-\epsilon_{+}(\k+\q)},\label{M}
\end{align}
where the mixing angle is
\begin{align}
\cos\phi_\k=\sqrt{\frac{\epsilon_+(\k)-\epsilon_{0,\k+\k_1}+\Delta_1}{\epsilon_+(\k)-\epsilon_-(\k)}},
\end{align}
and the eigenenergies from the unperturbed Hamiltonian in the EIT setup ($\Omega_1=0$) is 
\begin{align}
\epsilon_\pm(\k)=&-\Delta_1+\frac{\bar{\Delta}_2+\epsilon_{0,\k+\k_1}}{2}\nonumber\\&
\pm\frac{\sqrt{(\bar{\Delta}_2-\epsilon_{0,\k+\k_1})^2+4\Omega_2^2}}{2}.
\end{align}
The recoil-energy shifted detuning is $\bar{\Delta}_2\equiv\Delta_2+\epsilon_{0,\k+\k_r}$, the kinetic energy is $\epsilon_{0,\k}\equiv\k^2/(2m)-\mu$, and the recoil momentum is $\k_r\equiv\k_1-\k_2$.\ The chemical potential is $\mu$, and $\k_{1,2}$ are the central momenta of the probe and control fields respectively.\ A phenomenonological spontaneous decay rate ($\Gamma$) of the excited state can be added by replacing $\epsilon_{0,\k+\k_1}$ with $\epsilon_{0,\k+\k_1}-i\Gamma$.\ Note that we assume negligible dephasing rate between the two hyperfine ground states.

The above Eq. (\ref{chi}) provides the crucial recipe of the many-body effects on the EIT spectrum for generally any strongly correlated ultracold quantum gases \cite{Jen}.\ The nontrivial power-law dependence of the EIT spectrum near the resonance is shown for a Luttinger liquid, and the significant frequency shift and asymmetric absorption spectrum can be identified for a Bose-Mott insulator phase \cite{Jen}.\ This non-destructive EIT measurement is proposed to detect the Fermi paring in a Bardeen-Cooper-Schrieffer (BCS) superfluid state of two-component Fermi gases \cite{pairing}.\ The gap energy can be also measured from the transparency position \cite{Jen}.\ As long as the single-particle Green's function is known, the EIT spectrum can be derived.\ On the other hand, the EIT spectroscopy can be an efficient method to extract the single-particle Green's function.


The functional $M(\k+\q,\omega-\tilde{\omega})$ incorporates all the information of the laser parameters in the EIT setup, which is well defined.\ We find that Eq. (\ref{chi}) has a simple form of the convolution between the single-particle Green's function and $M$ in momentum-frequency space.\ Since the dispersion of the probe field has $\omega=q_z c$ in the propagating direction, the momentum (position) of $q_z$ ($z$) is not able to be resolved independently from frequency (time) in the EIT spectroscopy.\ We let $q_z$ $=$ $0$ in Eq. (\ref{chi}) which is valid if the spread of frequency $\omega$ $\ll$ $\omega_1$ so $\chi(q_z=|\omega|/c)$ $\approx$ $\chi(q_z=0)$.\ We first Fourier transform Eq. (\ref{chi}) in frequency space,

\begin{align}
\bar{\chi}(\bar{\q},t)=-\frac{2\pi d_0}{V}\sum_\k i\bar{G}^<(\k,t)\bar{M}(\k+\bar{q},t),\label{Green1}
\end{align}
where $\bar{\chi}(\bar{\q},t)$ and $\bar{M}(\k+\bar{\q},t)$ are the inverse Fourier transforms ($F^{-1}$) of $\chi(\bar{\q},\omega)$ and $M(\k+\bar{\q},\omega)$ respectively with $\bar{\q}$ $\equiv$ $(q_x,q_y)$.\ We may proceed to Fourier transform the momentum degree of freedom ($\bar{\q}$) above, and derive directly the transverse space and real-time Green's function $G^{<}(0,0;\bar{\r},t)$ in terms of $\bar{\chi}$ and $\bar{M}$ where $\bar{\r}$ $\equiv$ $(x,y)$.\ The validity of the Fourier transform in momentum space requires large enough $\bar{\q}$ to cover the complete EIT spectrum $\bar{\chi}(\bar{\q},t)$.\ To estimate the largest possible transverse range of the probe field momentum, it is limited by $1/\Delta L_{x,y}$ where $\Delta L_{x,y}$ is the average transverse atomic distance.\ In this limit however the probe field suffers from small optical density if the interacting cross section is less than $\Delta L_{x}\Delta L_{y}$.\ Therefore the probe field can not probe large enough momentum for general quantum degenerate gases (with the order of $\Delta L_{x,y}\sim 200$ nm for a typical density $\rho=10^{14}$ cm$^{-3}$), and we can not further Fourier transform Eq. (5).\ For simplification and to proceed the investigation of Green's function retrieval, we choose $\bar{q}=0$ and Eq. (\ref{Green1}) becomes

\begin{align}
\bar{\chi}(0,t)=-\frac{2\pi d_0}{V}\sum_\k i\bar{G}^<(\k,t)\bar{M}(\k,t),\label{Green}
\end{align}
where due to the limited spatial bandwidth of the EIT, we may not resolve directly the transverse dynamical Green's function from the EIT susceptibility $\bar{\chi}(0,t)$.\ 

However we will show in the examples of the next section that we may still extract the dynamical Green's function $\bar{G}^<(\k,t)$ from the information of $\bar{\chi}(0,t)$.\ The EIT spectrum holds much useful information to reconstruct $\bar{G}^<(\k,t)$ to provide the features that characterize the many-body systems.\ In addition we note that choosing a different $\bar{\q}$ in Eq. (\ref{Green1}) does not improve our reconstruction steps in the examples of section III, and in general it makes no significant difference than Eq. (\ref{Green}).

\subsection{Experimental setup}

In Fig. \ref{fig1}, we demonstrate an experimental setup to extract the single-particle Green's function using EIT spectroscopy.\ We use a spatio-temporal pulse to probe the ultracold quantum gas.\ The probe pulse goes through a 50:50 beam splitter (B.S.) to interact with the atoms, and the reflected one acts as a reference pulse.\ The transmitted probe pulse is then guided by the fiber to interfere with the reference one.\ We denote the process as the interferometry box that characterizes the temporal information of the transmitted pulse.\ This technique of measuring the spatial and temporal electric field \cite{pulse1, pulse2} is not new for the ultrafast optical society.\ The single trace of interferometry is done through the input fiber at the specific position $(x,y,z)$ that the electric field $E(t)$ can be resolved \cite{pulse3}.\ To further resolve the complete spatiotemporal information of the electric field pulse $E(x,y,z,t)$, the positions of the fiber is scanned with a high resolution of sub-$\mu$m \cite{pulse1}.\ 

The temporal information of the electric field can be characterized by interfering two input pulses, the reference [$E_{\rm ref}(\omega)$] and unknown [$E_{\rm unk}(\omega)$] ones respectively, into the interferometry box as shown in Fig. \ref{fig1}.\ Inside the box, these two pulses are collimated, and they cross at the camera with an angle $\theta=2d/f$ which is defined by the separation of two fibers with a distance of $2d$ and the focal length of the collimating lens $f$.\ We choose $\hat{z}$ as the propagation axis and the pulses cross in the axis $\hat{x}$.\ The oscillation part of the interfered spectrum becomes \cite{pulse3} 

\begin{align}
S(\omega,x)&=|E_{\rm ref}(\omega)||E_{\rm unk}(\omega)|\nonumber\\
&\times \cos\left(\varphi_{\rm ref}(\omega)-\varphi_{\rm unk}(\omega)+k_x x\sin\theta\right),\nonumber
\end{align}
where a diffraction grating is used to map the wavelengths (or frequency $\omega$) of pulses to the horizontal positions orthogonal to $\hat{x}$ and $\hat{z}$.\ The wavevectors of the collimated pulses are $\k=(\pm k_x\sin\theta\hat{x},k_z\cos\theta\hat{z})$, and the phases of the electric fields are $\varphi_{\rm ref}(\omega)$ and $\varphi_{\rm unk}(\omega)$.\ Since the unknown phase information can be inferred from the interference fringes in $\hat{x}$ axis, the amplitude and the phase of unknown electric field are respectively characterized by dividing out $|E_{\rm ref}(\omega)|$ and shifting $\varphi_{\rm ref}(\omega)$.

With the electric field interferometry, we extract the information of the transmitted pulse $\tilde{\Omega}'_1(0,\omega)$.\ The linear susceptibility $\chi(0,\omega)$ can be derived from $\tilde{\Omega}'_1/\tilde{\Omega}_1$ $=$ ${\rm exp}(i\k_1\chi L/(2\epsilon_0))$ in the paraxial approximation where $L$ is the propagation length.\ We then retrieve the single-particle Green's function according to Eq. (\ref{Green}), and the reconstruction steps are detailed in the next section.\ For the purpose of extracting the dynamical Green's function in the strongly correlated quantum gases, the temporal range of the probe pulse requires less than the natural lifetime of the probe field transition.\ We choose the low-lying Rydberg transition in EIT measurements to have significant many-body effects from the quantum gases \cite{Jen,Jen2}.\ The temporal range is then required to be the inverse of the EIT spectrum spread [$\sim 1/(4\Omega_2)$].\ It is in the order of $1/\Gamma$ where $\Gamma$ is several kHz for rubidium atoms with a low-lying Rydberg transition $|n'\rm{P}_{3/2}\rangle$ ($n'$ $\approx$ $20$) \cite{Rydberg}.\ The $\mu$s pulse is sufficient to probe the entire dynamical EIT spectrum in frequency space.\

Note that here we consider a homogeneous system such that the paraxial approximation is valid.\ In general the Maxwell-Bloch equation for the probe Rabi frequency ($\Omega_1$) in our EIT setup reads \cite{QO:Scully}
\begin{align}
\left[\frac{1}{c}\frac{\partial}{\partial t}+\frac{\partial}{\partial z} -\frac{i}{2\k_1}\nabla^2_\perp\right]\Omega_1(\r,t)=\frac{i\k_1}{2\epsilon_0}P(\r,t),\label{MB}
\end{align}
where the polarization operator $P(\r,t)$ can be derived from the linear perturbation of $\Omega_1$ as in Eq. (\ref{chi}).\ Therefore Eq. (\ref{MB}) becomes a self-consistent propagating equation for the probe field.\ The diffraction of the probe field is described by the Laplacian term ($\nabla^2_\perp$) which is crucial for an inhomogeneous system.\ Furthermore, the diffraction of the probe field has to be considered if the transverse distribution of the electric field is on the order of the central wavelength ($\sim 1/\k_1$).\ However in this case the small area of the light-matter interactions may invalidate Eq. (\ref{MB}) because of the requirement for large ensemble average over the cross section along the propagation \cite{quantization}.\ For the typical 3D Mott-insulator \cite{Bloch} we will study in the next section, the extent of the system is approximately 65 lattice sites ($\sim 26~\mu$m).\ Based on this length scale, the diffraction of the probe field is negligible for $\Delta\k\ll\k_1$ where $\Delta\k$ is the full-width of half-maximum (FWHM) of a transform-limit Gaussian pulse.


The longitudinal density distribution will indicate a spatial dependence for the electric susceptibility in Eq. (\ref{chi}).\ If we average out the spatial dependence, we expect the qualitatively same EIT spectrum as in a uniform gas, except for an extra inhomogeneous broadening in the spectrum due to the momentum/density distribution \cite{Jen}.\ We note that in the light propagation dynamics, the edge effect appears in the EIT setup in BEC \cite{BEC}.\ The spin wave inside the BEC will occupy near the cloud edge with a low density, which affects the fidelity of the storage process of the probe pulse \cite{BEC}.\ As in our case for the EIT spectroscopy, though the finite size effect influences the overall efficiency, it does not significantly modify the structure of the dynamical response of the transmitted probe field.

\section{Example: deep 3D Mott-insulator state and 1D Luttinger liquid}

In this section we take an example of a deep Mott-insulator (MI) state to demonstrate the finite temperature Green's function calculation.\ Then we investigate the reconstruction of dynamical Green's function based on Eq. (\ref{Green}) for MI and Luttinger liquid.

\subsection{Finite temperature single-particle Green's function}

In practice, the finite temperature effect should be taken into account in the ultracold quantum gases.\ We follow the formalism of linear response theory at finite temperature \cite{manyparticle}, and find that the EIT spectrum depends on the single-particle temperature Green's function,
\begin{align}
i\mathcal{G}^<(\r,t)={\rm Tr}\{\hat{\rho}_G\hat{\psi}^\dagger(\r,t)\hat{\psi}(0)\}, \label{temp}
\end{align}
where $\hat{\rho}_G$ $\equiv$ ${\rm exp}(-\beta\hat{K})$$/$${\rm Tr}({\rm exp}(-\beta\hat{K}))$ is the density operator for the grand canonical ensemble, and $\hat{\psi}(\r,t)$ is the ground state operator for the Hamiltonian $\hat{K}=\hat{H}-\mu\hat{N}$.\ $\hat{H}$ is the unperturbed Hamiltonian for the EIT setup without the probe light, and $\beta=1/(k_B T)$ where $k_B$ is the Boltzmann constant, and $T$ is the temperature of the gas.\ The electric susceptibility of EIT at finite temperature is then similar to Eq. (\ref{chi}) with the replacement of the single-particle Green's function by a finite temperature one in Eq. (\ref{temp}).\

To derive the single-particle temperature Green's function, we may utilize the spectral function $A(\k,\omega)$ \cite{Mahan} that
\begin{align}
i\tilde{\mathcal{G}}^<(\k,\omega)&=\mp f_{F/B}(\omega)A(\k,\omega), \nonumber\\
A(\k,\omega)&=-2{\rm Im}[\tilde{\mathcal{G}}_{\rm ret}(\k,\omega)],
\end{align}
where $f_{F/B}(\omega)$ is the Fermi or Boson number distributions at finite temperature $T$, and $\mathcal{G}_{\rm ret}$ is the retarded Green's function.

As an example we consider the MI of strongly interacting bosons in a 3D optical lattice.\ In a single band Hubbard model (HM) \cite{BHM}, the ground state field operator is $\hat{\psi}_g(\r,t)$ $=$ $\sum_\mathbf{R} \hat{g}_\mathbf{R}(t) w_\mathbf{R}(\r)$ where $w_\mathbf{R}(\r)$ is the Wannier function, and $\hat{g}_\mathbf{R}(t)$ is the field operator at site ${\bf R}$.\ When deep inside the MI state, we use the three-state model \cite{three,Mott} to obtain the Green's function at finite temperature in the limit of large on-site interaction $U$,
\begin{align}
i\bar{\mathcal{G}}^<(\k,t)&=\sum_{\mathbf{R}}|\tilde{w}_{\mathbf{R}}(\k)|^2e^{-iUt/2}\theta(t)\left[f_B\left(\frac{U}{2}\right)(n_0+1)\right.\nonumber\\
&\left. +\left(1+f_B\left(\frac{U}{2}\right)\right)n_0\right],\label{Mott_temp}
\end{align}
where for convenience we calculate the real time single-particle temperature Green's function in momentum space.\ $\theta(t)$ is a step function, $n_0$ is the integer filling fraction, and $\tilde{w}_{\mathbf{R}}(\k)$ is the Fourier transform of $w_\mathbf{R}(\r)$.\ Note that the particle and hole excitation energies $\epsilon_{p(h)}(\k)$ $\approx$ $U/2$ when deep inside the Mott state.\ 

The first and second parts in the bracket of Eq. (\ref{Mott_temp}) represent the contributions from the particle and hole excitations respectively.\ At zero temperature, only the hole excitation remains as the quantum many-body effect on the EIT spectrum \cite{Jen}.\ In the next subsection, we demonstrate the reconstruction of dynamical Green's function in the example of 3D MI via EIT spectroscopy.

\subsection{Reconstruction of Green's function in MI}

First we calculate the electric susceptibility in Eq. (\ref{chi}) from Eq. (\ref{Mott_temp}), and we have
\begin{align}
\chi_{\rm MI}(\q,\omega)&
=\frac{d_0N}{V}\sum_\k|\tilde{w}_\mathbf{R}(\k)|^2\bigg[\frac{\cos^2\phi_{\k+\q}}{\omega+U/2+\epsilon_{-}(\k+\q)}\nonumber\\
&+\frac{\sin^2\phi_{\k+\q}}{\omega+U/2+\epsilon_{+}(\k+\q)}\bigg]\left(1+3f_B\left(\frac{U}{2}\right)\right),
\label{Motteq}
\end{align}
where we consider a unit filling phase, $n_0=1$.\ The finite temperature effect on the MI is the fluctuation of the total number of particles, so $i\bar{\mathcal{G}}^<(\k,t)$ is not much different from $i\bar{G}^<(\k,t)$ in the deep Mott limit.

In practical EIT experiments, because of the restriction on finite lattice spacing $d=426$ nm, we may not probe the dynamical response at momentum higher than $\sim 1/d$.\ This essentially limits the spatial bandwidth the probe field could provide.\ Therefore with this limit and considering of the transverse length in the experiment \cite{Bloch}, we let $\q=0$, and from Eq. (\ref{Motteq}) we derive $\bar{\chi}_{\rm MI}(0,t)$ (for simplicity at zero temperature),
\begin{align}
\bar{\chi}_{\rm MI}(0,t)=\sum_\k|\tilde{w}_\mathbf{R}(\k)|^2 e^{-iUt/2}\bar{M}(\k,t),\label{Motteq_t}
\end{align}
where we normalize the above by the constants of $d_0$ and atomic density, and
\begin{align}
\bar{M}(\k,t)=\theta(t)[\cos^2\phi_{\k}e^{-i\epsilon_{-}(\k)t} + \sin^2\phi_{\k}e^{-i\epsilon_{+}(\k)t}],\label{Mt}
\end{align}
where $\theta(t)$ is a step function. Note that $\bar{\chi}_{\rm MI}(0,t=0)$$=$$1$.\

\begin{figure}[t]
\centering\includegraphics[height=4cm, width=8cm]{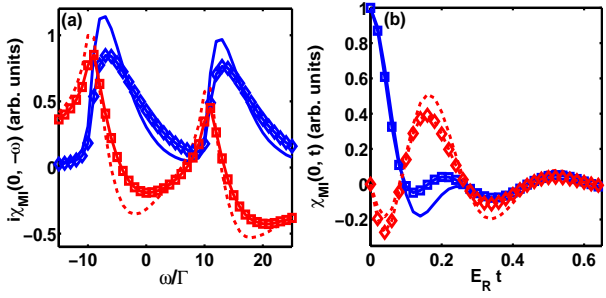}
\caption{(Color online) The EIT profiles of a 3D unit-filling Mott-insulator with $^{87}$Rb atoms (lattice constant $d=426$ nm) in the counterpropagating excitation scheme.\ We use a resonant control field ($\Delta_2=0$) with Rabi frequency $\Omega_2=10~\Gamma$, and set $\Delta_1=0$.\ The excited state is chosen as low-lying Rydberg transition of $|24\textrm{P}_{3/2}\rangle$ with the spontaneous decay rate, $\Gamma^{-1}=28.3~\mu\text{s}$.\ The corresponding parameters ($U$, $J$) in unit of the recoil energy $E_R$ are (0.5, 0.007) and (1.05, $10^{-4}$) for $V_0/E_R=15$ (solid-blue and dash-red) and $40$ ($\Diamond$ and $\square$) respectively \cite{OL}.\ (a) Absorption (Re[$-i\chi_{\rm MI}$], solid-blue, $\Diamond$) and dispersion (Im[$i\chi_{\rm MI}$], dash-red, $\square$) profiles are demonstrated.\ We plot $i\chi_{\rm MI}$ versus $-\omega$ to demonstrate a normal dispersion inside the transparency window for comparisons with conventional EIT.\ (b)EIT spectrum $\bar{\chi}_{\rm MI}(0,t)$ in time domain with its real (solid-blue, $\square$) and imaginary (dash-red, $\Diamond$) parts.}%
\label{fig2}
\end{figure}

In Fig. \ref{fig2}, we show the dynamical EIT spectrum in frequency and time domains.\ An inhomogeneous broadening is clear in Fig. \ref{fig2}(a) when a deeper MI is investigated \cite{Jen}.\ In Fig. \ref{fig2}(b), we plot the EIT spectroscopy in time domain, and find the essential information of momentum width ($\sigma$) in the Wannier function and gap order parameter ($U$) can be found from the first and second derivatives of $\bar{\chi}_{\rm MI}(0,t)$,
\begin{align}
\frac{d}{dt}\bar{\chi}_{\rm MI}(t)|_{t=0}&=-i\frac{U}{2}-\hbar\Gamma-i\frac{\hbar^2\k_1^2}{2m}-i\frac{3\hbar^2}{2m\sigma^2},\label{eq1}\\
\frac{d^2}{dt^2}\bar{\chi}_{\rm MI}(t)|_{t=0}&=-iU\frac{d}{dt}\bar{\chi}_{\rm MI}(0)+\frac{U^2}{4}-B\nonumber\\
&-\big[\frac{15}{4\sigma^4}(\frac{\hbar^2}{m})^2-i\hbar\Gamma\frac{3\hbar^2}{m\sigma^2}+\frac{\hbar^2\k_1^2}{2m}\frac{5\hbar^2}{m\sigma^2}\big],\label{eq2}
\end{align}
where we assume a Gaussian function of width ($\sigma$) for the single band Wannier function, and the constant $B$ is
\begin{align}
B=\hbar^2(\Omega_2^2-\Gamma^2)+\frac{\hbar^2\k_1^2}{2m}(\frac{\hbar^2\k_1^2}{2m}-i2\Gamma).\nonumber
\end{align}

\begin{figure}[t]
\centering\includegraphics[height=4.5cm, width=8cm]{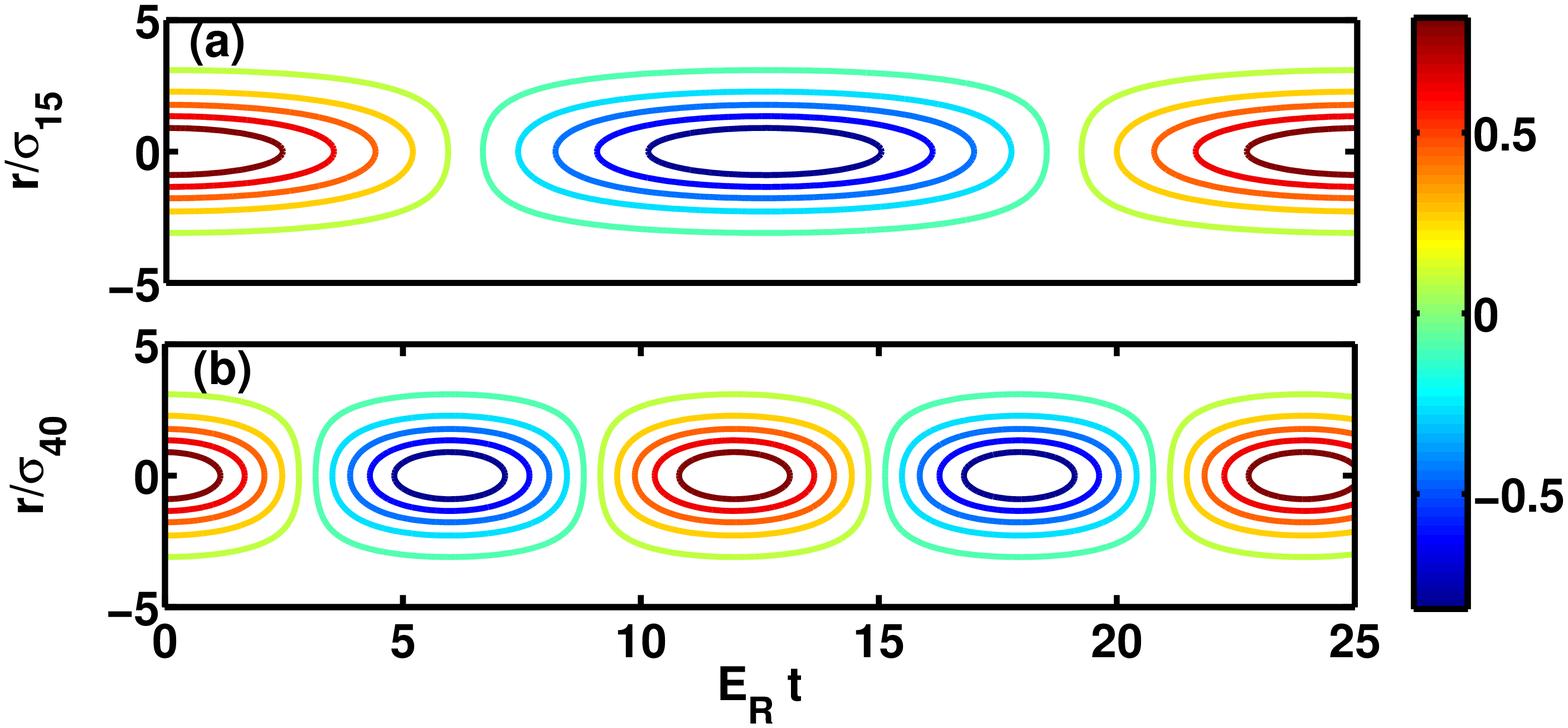}
\caption{(Color online) The real part of the single-particle Green's function [$iG^<(\bar{\r},t)$] for a unit-filling 3D Mott insulator (MI).\ The on-site interaction $U$ in units of the recoil energy $E_R$ ($\approx$ $20$ $k$Hz) of the trapping lasers are $0.5$ and $1.05$ respectively for (a) $V_0/E_R$ $=$ $15$ and (b) $40$.\ The oscillation period of the Green's function determines the inverse of half the gap order parameter in a MI.\ The spatial distribution of Green's function is plotted as $\bar{r}=\sqrt{x^2+y^2}$ where we use a Gaussian distribution as an approximate single band Wannier function $w_\mathbf{R}(\r)$, and $\sigma_{15,40}$ are 69 and 54 nm respectively.}%
\label{fig3}
\end{figure}

From Eqs. (\ref{eq1}) and (\ref{eq2}), we can solve for the unknown values of $\sigma$ and $U$, and we can reconstruct the dynamical Green's function of MI as
\begin{align}
&iG^{<}(0,0;\bar{\r},t)\propto e^{-\r^2/(4\sigma)}e^{-iUt/2},
\end{align}
where we show in Fig. \ref{fig3}.\ The real-time dynamics of Green's function indicates a modulation of the period which is the inverse of the on-site interaction energy.\ Therefore the EIT spectroscopy along with Green's function reconstruction can be the alternative approach to lattice potential gradient \cite{Bloch} or modulation spectroscopy \cite{Esslinger} to determine the gap energy in MI.\ 

The presumption of single band Wannier function and gap order parameter are crucial for reconstructing the Green's function of a 3D MI.\ Since we assume only two parameters for Green's function, it is not surprising that the information of the slopes at $\bar{\chi}_{\rm MI}(0,t=0)$ in Eqs. (\ref{eq1}) and (\ref{eq2}) infers the essential parameters, and they can be expressed as
\begin{align}
\frac{d}{dt}\bar{\chi}_{\rm MI}(0,t)|_{t=0}&=\int i\omega\chi_{\rm MI}(0,\omega)d\omega=i\langle\omega\rangle,\\
\frac{d^2}{dt^2}\bar{\chi}_{\rm MI}(0,t)|_{t=0}&=-\int\omega^2\chi_{\rm MI}(0,\omega)d\omega=-\langle\omega^2\rangle,
\end{align}
where $\int\chi_{\rm MI}(0,\omega)d\omega$$\equiv$$\bar{\chi}_{\rm MI}(0,t=0)$ which is normalized.\ From the above we show how to extract the parameters of Green's function, which is equivalent to find the average of the frequency and its variance (second order moment) for a normalized EIT spectrum.\ The Green's function information embeds in the EIT spectroscopy in which we may need higher order moments to extract more complicated forms of Green's functions.
\subsection{Reconstruction of Green's function in Luttinger liquid}

For the second example, we demonstrate the single-particle Green's function of the Luttinger liquid (LL) which is a universal 1D effective model \cite{Haldane1, Haldane2, LL}.\ The single-particle Green's function can be exactly calculated by the Bosonization method \cite{Giamarchi,LL}, which is 
\begin{align}
iG^<_{\rm LL}(x,t) =\frac{na^{1/(2\kappa)}}{[x^2+(a+ivt)^2]^{1/(4\kappa)}}.\label{LL}
\end{align}
The Luttinger parameter is $\kappa$, $v$ is the phonon velocity, $a^{-1}$ is the system-dependent momentum cutoff, and $n$ is the atomic density.\ $1$$<$$\kappa$$<$$\infty$ is for a short-range repulsive interaction, while $\kappa$ can be smaller than one if the interaction is long ranged.

\begin{figure}[t]
\centering\includegraphics[height=4cm, width=8cm]{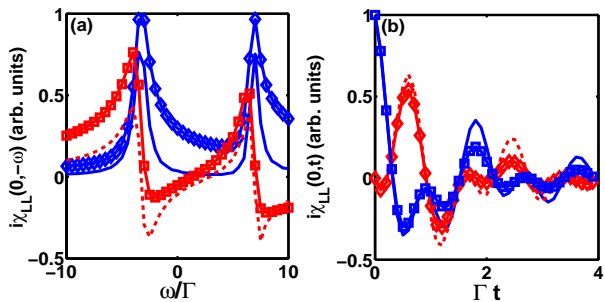}
\caption{(Color online) The EIT profiles for a Luttinger liquid of $^{87}$Rb atoms.\ The driving conditions are the same as in Fig. \ref{fig2} but with Rabi frequency $\Omega_2=5~\Gamma$.\ (a) Absorption (Re[$-i\chi_{\rm LL}$], $\Diamond$, solid-blue) and dispersion (Im[$i\chi_{\rm LL}$], $\square$, dash-red) profiles are demonstrated respectively for $\kappa$$=$$1,10$.\ (b) EIT spectrum $\bar{\chi}_{\rm LL}(0,t)$ in time domain with its real (solid-blue, $\square$) and imaginary (dash-red, $\Diamond$) parts.\ For a typical experimental regime \cite{1D}, we may take the phonon velocity $v$$=$$4.3$ mm/s and the cutoff $a$$=$$0.12$ $\mu$m.}%
\label{fig4}
\end{figure}

The electric susceptibility can be also derived by substituting Eq. (\ref{LL}) into Eq. (\ref{chi}) \cite{Jen}.\ We may proceed to find $\bar{\chi}_{\rm LL}(0,t)$ from Eq. (\ref{Green}) as
\begin{align}
\bar{\chi}_{\rm LL}(0,t)=\sum_\k \int \frac{a^{1/(2\kappa)}e^{-i\k x}dx}{[x^2+(a+ivt)^2]^{1/(4\kappa)}}\bar{M}(\k,t),
\end{align}
which is normalized, and reconstruct the dynamical Green's function from its first order derivative at $t=0$,
\begin{align}
\frac{d}{dt}\bar{\chi}_{\rm LL}(0,t)|_{t=0}=-\frac{iv}{2\kappa a}-\hbar\Gamma-i\frac{\hbar^2\k_1^2}{2m}-i\frac{\sqrt{2}\hbar^2}{4ma^2\kappa},
\end{align}
where we extract the Luttinger parameter $\kappa$ from the measurement of the first order moment $i\langle\omega\rangle$ in EIT spectroscopy.\ Note that we have assumed the only unknown parameter in Green's function of 1D quantum gas is $\kappa$, and in general we may extract for example the phonon velocity or system-dependent cut-off length scale by measuring higher order moments in EIT spectroscopy.\ In Fig. \ref{fig4} we show the dynamical EIT spectrum in frequency and time domains.\ The inhomogeneous broadening and shifted transparency position are due to the strong interactions. 

\begin{figure}[t]
\centering\includegraphics[height=5cm, width=8.25cm]{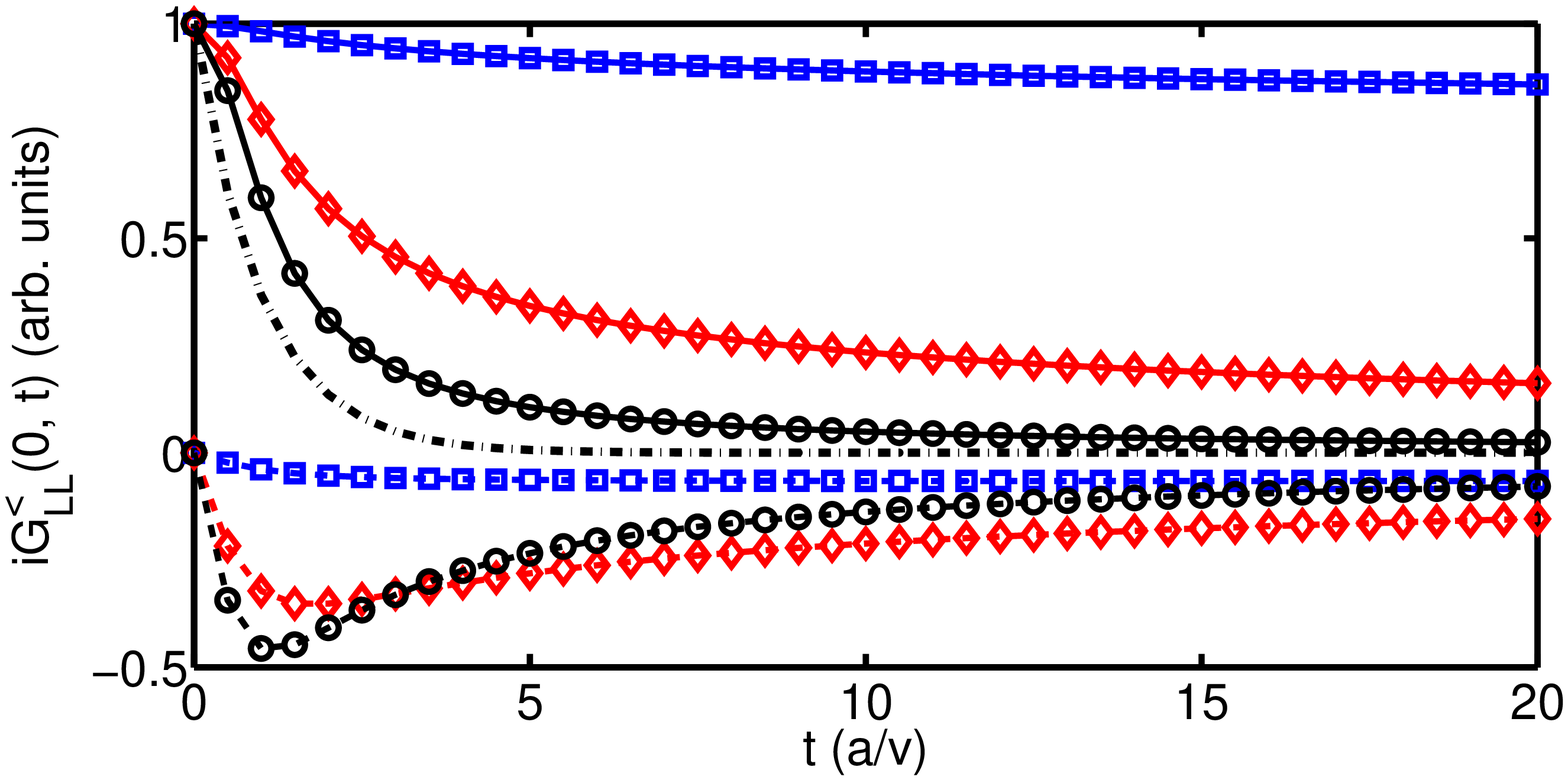}
\caption{(Color online) The single-particle Green's function [$iG_{\rm LL}^<(0,t)$] for a Luttinger liquid.\ The Luttinger parameters are $\kappa$ $=$ $0.6$ ($\circ$), $1$ ($\Diamond$), $10$ ($\square$) for the real (solid) and imaginary (dash) parts of $iG_{\rm LL}^<(0,t)$.\ Long-range order is present for a weakly interacting gas ($\kappa$ $=$ $10$) while a strongly interacting gas with $\kappa$ $=$ $0.6$ or $1$ shows a power-law decay by time.\ The dash-dot line is $\exp(-t)$ to compare with a power-law decay.}%
\label{fig5}
\end{figure}

In Fig. \ref{fig5}, we show the real-time single-particle Green's function of LL.\ For a weakly interacting quantum gas ($\kappa=10$), the relatively flat correlation function indicates the long-range order in the superfluid.\ The strongly interacting quantum gas ($\kappa=0.6,1$) on the other hand shows a noninteger power-law decay in the Green's function.\  This quasi-long-range order originates from the collective excitations in a LL model with a linear dispersion $\omega(\q)=|\q|v$ in the low-temperature and long-wavelength limit.\ In our proposed experiment, the EIT spectroscopy provides a nondestructive way to extract the LL parameter $\kappa$ which is essential to characterize the 1D bosonic quantum gas.

\section{Discussion and Summary}

Our proposed EIT spectroscopy may be also applied to the repulsive 1D two-component fermions.\ It can provide a genuine observation of the two-fold linear excitations in the spin and the charge sectors from the extraction of the single-particle Green's function \cite{manybody} similar to our case of a bosonic LL.\ For attractive 1D spin-$1/2$ fermions, the spin gap order parameter is also observable in the EIT spectroscopy in analogy to the BCS superfluid of two-component Fermi gases \cite{Jen, pairing}.\ Therefore, the EIT spectroscopy can be applied to investigate on a plethora of low-dimensional quantum many-body systems with arbitrary interactions.

As a comparison with the method of Ramsey interferometry to extract the spin correlation functions \cite{Knap}, EIT spectroscopy provides a non-destructive probe of the many-body system while the interferometry involves the destructive projection measurement in the Ramsey pulse sequence.\ Valuable information can be extracted from the spin correlation function to determine the quantum phase transition from the characteristic scaling behavior \cite{Knap}.\ Similarly, EIT measurement extracts the important scaling factor of the power-law decay function for example in 1D LL.\ A common difficulty in practice is the limited bandwidth in experiments which in our case restricts the direct probe of the transverse spatial dynamics of Green's function in the ultracold atomic gases.\ However, we may still deduce the Green's function from the post-calculation of the frequency moments in the EIT spectroscopy.

In summary, we show that the single-particle Green's function can be extracted by an efficient and nondestructive EIT spectroscopy.\ We propose to utilize an electric field interferometry to characterize the temporal profiles of the probe fields.\ From the information of the transmitted probe pulse or equivalently the EIT dynamical response, we may deduce the essential single-particle Green's function of the ultracold quantum gases.\ In the examples we take 3D Mott insulator and 1D LL to demonstrate that the gap order and LL parameters respectively can be extracted from the information of EIT spectrum.\ This provides an all optical approach to reveal the essential single-particle Green's function in the ultracold quantum gases, and opens up a new avenue to study many-body physics in the strongly correlated system.\ 

\section*{ACKNOWLEDGMENTS}
We appreciate fruitful discussions with T. Pohl and Ite A. Yu.\ We also thank M. Knap and E. Demler for pointing out the relevant Ramsey interferometry method to probe the correlation functions \cite{Knap}.\ This work is supported by NSC and NCTS grants in Taiwan.


\end{document}